\documentclass{article}
\usepackage{graphicx}

\begin{document}

\begin{center}
{\Large\bf ON THE GRAVITATIONAL FIELD OF A POINT MASS IN EINSTEIN UNIVERSE BACKGROUND}
\\[1.5cm]
Dumitru Astefanesei
\\
{\small \it Department of Physics, McGill University, Montreal, QC, H3A 2T8, Canada
\\
e-mail: astefand@physics.mcgill.ca
\\(on leave from Al. I. Cuza University, Blvd. Copou 11, Iasi, Romania)} 
\\[0.5cm]
and 
\\[0.5cm]
Eugen Radu
\\
{\small \it Albert-Ludwigs-Universit\"at Freiburg, Fakult\"at f\"ur Physik, 
\\
\it Hermann-Herder-Stra\ss e 3, D-79104 Freiburg, Germany
\\ email: radu@newton.physik.uni-freiburg.de
\\(on leave from Institute of Technical Physics, D. Mangeron 47, Iasi, Romania)}
\\[1cm]
\end{center}
\begin{abstract}

Some properties of an exact solution due to Vaidya, 
describing the gravitational 
field produced by a point particle in the 
background of the static Einstein universe are examined. 
The maximal analytic extension and the nature of 
the singularities of the model are discussed.
By using the Euclidean approach, some quantum aspects are analysed
 and the thermodynamics of this spacetime is also discussed.

\end{abstract} 

\newpage

\section{Introduction}
The standard Schwarzschild solution is described in flat background. 
Because a black hole is a cosmological object, it is 
worthwhile to examine the effect of the cosmological background 
on the black hole properties. We recall that by 
considering a non-zero cosmological constant we obtain the Schwarzschild-
(anti)de Sitter solution with rather different properties.
Also, we would like to take into account the fact that a black hole may  
also be surrounded by a local mass distribution.

A model taking into account the deviation from flatness on a large scale was 
proposed a long time ago by Einstein and 
Strauss \cite{1}. 
In this model, the vacuum Schwarzschild field matches accross a spherical comoving boundary
to a pressure-free Friedmann-Robertson-Walker (FRW) universe.
This would provide for example a description of the effect of the cosmic expansion 
on the gravitational field of the Sun.

Another way to deal with the embedding of massive objects in a 
FRW universe is to solve 
Einstein's field equations exactly or approximately,
 in such a way that the resulting solution can be interpreted 
as an embedding of some massive object in the considered background. 
In 1933, McVittie found solutions of Einstein's 
field equations for a perfect fluid energy-tensor, 
representing a Schwarzschild field embedded in the FRW spacetime \cite{2}
(see also ref. \cite{3, 4, 5}  for an up-to-date discusion of the 
Mc-Vittie solutions' properties ).

A rather different approach has been considered by Vaidya, who derived a perfect fluid solution, 
interpreted as the Kerr metric in the cosmological background of the 
Einstein space \cite{6, 7}.

When specializing the Vaidya solution  
for a vanishing angular momentum, we obtain the simpler case of a 
Schwarzschild metric in the Einstein space background.
 
It is the purpose of this paper to study some properties of 
this spacetime, in an attempt to address the fundamental question of how
the cosmological background will affect the black hole properties.
    
The paper is structured as follows:  in section 2 we outline the basic features of 
the Vaidya solution,
while in section 3 we analyse the global properties of this spacetime
 and the issue of singularities of the model.
Based on these results, some thermodynamic properties are discussed in section 4.
We conclude with section 5 where the results are compiled.

\section{Einstein equations and matter content}
In the particular case of zero angular momentum, 
the metric proposed by Vaidya reduces to the form  
\begin{equation}\label{metric}
ds^2 = \frac{dr^2}{1-\frac{2m}{R_{0}}\cot\frac{r}{R_0}} +
 R_0^2 \sin^2(\frac{r}{R_0})(d\theta^2 +
\sin^2\theta d\phi^2)-(1-\frac{2m}{R_0}\cot\frac{r}{R_0})dt^2.
\end{equation}
This metric is also a particular case of the Wahlquist solution \cite{8}. 
$R_0$ and $m$ are two positive constant; 
when setting $m$ equal to zero,
 the metric (\ref{metric}) reduces to the metric of the Einstein universe; 
for $R_0 \to \infty$ the vacuum Schwarzschild solution is recovered.
The parameter $R_0$ can be considered as a measure of the cosmological 
influence on the black hole properties.
Therefore it is natural to interpret $m$ and $R_0$ as representing the  
Schwarzschild mass and the radius of the universe, respectively.

It is also interesting that the only metric of the form
\begin{equation}
ds^2=e^{2A(r)}dr^2 +  R_0^2 \sin^2(\frac{r}{R_0})(d\theta^2 + \sin^2\theta d\phi^2)
- e^{-2A(r)}dt^2,
\end{equation}
satisfying the condition
\begin{equation}
R_r^r=R_\theta^\theta=R_\phi^\phi
\end{equation}
(with $R_a^b$ the Ricci tensor) is (\ref{metric})
 (such an universe will be isotropic around one observer, but not spatially 
homogeneous). 
When solving the Einstein equations, we find, 
in addition to the matter content 
of the Einstein universe, a perfect fluid
 with $\rho+3p=0$,
violating in some regions the energy conditions.
As opposed to McVittie's solution, this matter content 
is not spatially homogeneous, since
\begin{eqnarray}\label{ro}
8\pi\rho&=&\frac{6m}{R_0^3}\cot(\frac{r}{R_0}),
\nonumber\\
8\pi p&=&-\frac{2m}{R_0^3}\cot(\frac{r}{R_0}).
\end{eqnarray}
This new contribution to the total energy-momentum tensor, 
associated with the energy source at the origin of the radial coordinate,
 will spoil the homogeneous 
nature of the Einstein universe.
We can think about the above relations as describing 
the perturbation of the local matter content of the Einstein 
cosmological background induced by a point mass.
It is also tempting to suppose a quantum origin of this stress-energy tensor.


It is well known that the Einstein universe solution is unstable (see $e.g.$ \cite{12}). 
Also, in the median region $r \sim R_0\pi/2$, the local properties 
of the Vaidya model are well described by the Einstein universe.
However, a local perturbation of the cosmic
background in this region will propagate towards antipodes.
Therefore, the radius $R_0$ would continue to change in the same direction 
once the model started to expand or contract.
One may expect that the new time-dependent configuration 
will represent the static case of a general solution briefly noted in \cite{6} 
and interpreted as the Kerr field embedded in a FRW universe.

This raises the interesting question if this time-dependent 
solution has a central singularity and event horizon, 
and if so, how they are affected by the cosmic expansion.

These issues are currently being studied and will be discussed elsewhere.

We note also the form of the line-element (\ref{metric}) in Schwarzschild coordinates
\begin{eqnarray}\label{sch}
ds^2&=&\frac{d\overline{r}^2}{(1-(\frac{\overline{r}}{R_0})^2)(1-\frac{2m}{\overline{r}}
\sqrt{1-(\frac{\overline{r}}{R_0})^2})}+\overline{r}^2((d\theta^2 + \sin^2\theta d\phi^2)\nonumber\\
& &-(1-\frac{2m}{\overline{r}}\sqrt{1-(\frac{\overline{r}}{R_0})^2})dt^2,
\end{eqnarray}
where

\begin{equation}
\overline{r}=R_0\sin\frac{r}{R_0}.
\end{equation}

This helps us to compare with the results obtained in \cite{9}, 
where the gravitational field produced by a point mass in the 
background of the static Einstein universe is studied with many similar conclusions.
For small $m/R_0$, the line element (\ref{sch}) reproduces the approximate
solution derived in \cite{9}.

\section{Maximal analitic extension}
A metric of the form (2) has singularities where $e^{2A}$ vanishes or becomes infinite. 
However, some of these 
singularities can be pseudosingularities, caused by an inappropiate coordinate system. 
Following the usual techniques
of analytic extension across pseudosingularities \cite{10},
 we obtain Kruskal-like form of the metric (\ref{metric})
\begin{equation}
ds^2=\frac{32m^3
\exp(-\frac{r}{2m})}{[1+(\frac{2m}{R_0})^2]^2R_0\sin(\frac{r}{R_0})}(dz^2-dT^2)+
R_0^2\sin^2(\frac{r(z,T)}{R_0})(d\theta^2 + \sin^2\theta d\phi^2),
\end{equation}
which is related to the original form by
\begin{equation}\label{m1}
z^2-T^2=\frac{R_0}{2m} \exp(\frac{r}{2m}) 
\sin\frac{r}{R_0}(1-\frac{2m}{R_0}\cot\frac{r}{R_0}),
\end{equation}
\begin{equation}\label{m2}
\frac{T}{z}=\tanh \{\frac{t}{4m}[1+(\frac{2m}{R_0})^2]\}\qquad
r>r_H,
\end{equation}
\begin{equation}\label{m3}
\frac{z}{T}=\tanh \{\frac{t}{4m}[1+(\frac{2m}{R_0})^2]\}\qquad
r<r_H,
\end{equation}
where $r_H=R_{0} \arctan (\frac{2m}{R_0})$.

No singularities occur here, except the genuine singularities at $r=0$ 
and $r=\pi R_0$, which cannot 
be removed by any coordinate transformations.

On the manifold defined by coordinates $(z,\theta,\phi, T)$ with metric (8), 
we define four regions (fig.1): 
the region I with $z>|T|$ is isometric to the region of the metric
 (\ref{metric}) for which $r>r_H$. 
There is also a region I', defined by $z<-|T|$, 
which turns out to be again isometric with the same region 
of the metric (\ref{metric}). This can be regarded as another universe 
on the other side of the throat $r=r_H$. 
The regions II and II' are isometric with the region $r<r_H$ of the metric (\ref{metric}).
 The surface $r=r_H$ presents all the 
characteristics of an event horizon.
 For a freely falling observer an infinite time $t$ is required to traverse 
the finite distance $L_0$ between an exterior point and a point on the horizon, 
but that destination is reached in a finite proper time. 
A photon would require an infinitely long time to cover the finite stretch $L_0$. 
Once a particle has fallen inside $r=r_H$ it cannot avoid the singularity. 
In Schwarzschild coordinates, the horizon is located at 
$\overline{r}=\frac{2m}{\sqrt{1+(\frac{2m}{R_0})^2}}$; 
the effect of the curvature of the universe on the radius of 
the event horizon is to reduce it.
We note also that both $\rho$ and $p$ defined by (\ref{ro})
are well defined at $r=r_H$.

Both $r=0$ and $r=\pi R_0$ are curvature singularities with corresponding 
infinities of the functions $\rho$, $p$. 
However the $r=0$ singularity is less dangerous being located inside an event horizon. 
The situation with an antipodal singularity at $r=\pi R_0$ is rather different. Since there is no horizon, this is a naked singularity, contradicting the 
cosmic censorship conjecture (however, 
it is hazardous to claim that the Vaidya metric (\ref{metric}) describes the final state of a collapsing matter distribution). 
This singularity appears to be repulsive: no timelike geodesic hits them,
 though a radial null geodesic can. 
Then we can interpret the metric (\ref{metric}) as describing a black hole with an 
antipodal naked singularity in the 
cosmological background of Einstein universe.

In fact, given the high symmetry and the simple matter content of the solution we
 have to expect the existence of an 
antipodal singularity. 
However, we suspect that this is not
a generic property. We can hope that by considering a more general 
matter content this unpleasant feature can be 
avoided by the introducing an additional event horizon.
\section{Thermodynamic properties}
The presence of a naked singularity makes difficult the formulation of a quantum 
field theory on the background of 
metric (\ref{metric}). However, the standard arguments of the Euclidean approach for deriving 
the Hawking radiation seems to be 
valid in this case, too. We assume the possibility of extending path integral
 formulation of gravitational thermodynamics 
to the situation under consideration. This is a very strong assumption, given 
the existence of the naked singularity; 
we recall that one of the reasons to deal with the Euclideanization procedure 
was to avoid the space-time singularities 
\cite{11}.

Proceed by making in (\ref{m1}) the formal substitution $\xi=iT$ to yield
\begin{equation}
ds^2=\frac{32m^3 \exp(-\frac{r}{2m})}{[1+(\frac{2m}{R_0})^2]^2 R_0\sin(\frac{r}{R_0})}
(dz^2+d\xi^2)+R_0^2\sin^2(\frac{r(z,\xi)}{R_0})(d\theta^2 + \sin^2\theta d\phi^2).
\end{equation}

On the section on which $z,\xi$ are real, $r$ will be real and great or equal to $r_H$ 
(the singularity at $r=\pi R_0$ is still present). 
Define the imaginary time $\tau$ by $\tau=-it$. It follows from eq. (\ref{m2}) that
 $\tau$ is periodic with period
\begin{equation}\label{beta}
\beta=\frac{8\pi m}{1+(\frac{2m}{R_0})^2}.
\end{equation}
The periodicity in the imaginary (Euclidean) time is usually interpreted as 
evidence of a thermal bath of temperature 
$T=\frac{1}{\beta}$, so that the Hawking temperature of a Schwarzschild black hole 
in the Einstein universe background 
is identified as

\begin{equation}
T_H=\frac{1}{8\pi m}[1+(\frac{2m}{R_0})^2].
\end{equation}

The same result can be obtained by direct calculation of the surface gravity $k$ 
($T_H=\frac{k}{2\pi}$). 
The Hawking temperature of the system appears to be increased relative to that of 
Schwarzschild black hole of equal horizon
 area.

From the formula (\ref{beta}) one can see that $\beta$ has a maximum value $2\pi R_0$ 
for $m=R_{0}/2$ and therefore $T_H$ has a 
minimum value of $T_0 =(2\pi R_0)^{-1}$ when $r_H=\pi R_0/4$. 
For $m>\frac{R_0}{2}$, the temperature $T_H$ increases with the mass; 
for a large enough $m$ we have 
$T_H\approx\frac{m}{2\pi R^2_0}$.

According to $"Tolman$ $relation"$, the local temperature $T(r)$ measured by a moving, 
accelerated detector is related to 
the Hawking temperature $T_H$ measured by a detector in the asymptotic region by \cite{12}
\begin{equation}
T(r)=(-g_{\tau\tau})^{-\frac{1}{2}}T_H=\frac{1}{8\pi m}(1+
(\frac{2m}{R_0})^2)(1-\frac{2m}{R_0}\cot\frac{r}{R_0})^{-\frac{1}{2}}.
\end{equation}

We can consider this solution as describing a "dirty" black hole \cite{13} since 
the interaction with a classical matter 
field is present; 
however, the metric (\ref{metric}) is not asymptotically flat and the general formalism 
for analysing the 
thermodynamic properties of a dirty black hole cannot be applied \cite{13, 14}.

Accordingly to Gibbons and Hawking \cite{11}, thermodynamic functions including 
the entropy can be computed directily from 
the saddle point approximation to the gravitational partition function 
(namely the generating functional analytically 
continued to the Euclidean spacetime).

The Euclidean gravitational part of the action has the general form \cite{15}
\begin{equation}\label{action}
I_E=-\frac{1}{16\pi}\int_{M}(R-2\Lambda)\sqrt{g}d^4x + 
\frac{1}{8\pi}\int_{\partial M} K\sqrt{h}d^3x.
\end{equation}

 In this case, due to the global structure, it is not necessary a so called 
 "\textsl{reference action}" substraction 
(this substraction was needed for a Schwarzschild black hole in order to get 
 a finite result \cite{11}). In the semiclassical approximation, the dominant 
 contribution to the path integral will 
come from the neighborhood of saddle points of the action, that is, of classical 
solution; the zeroth order contribution to 
$\log Z$ will be $-I_E$.

The integral (\ref{action}) is evaluated on the Euclidean section (with $r\ge r_H$) 
containing the naked singularity at $r=\pi R_0$;
however it takes a finite value
\begin{equation}
I_E=\frac{\beta m}{2}[2\frac{1-(\frac {2m}{R_0})^2}{1+(\frac {2m}{R_0})^2}-
\frac{R_0}{2m} \arctan(\frac{2m}{R_0})+\beta \pi\frac{ R_0}{4}].
\end{equation}

 All thermodynamic properties can be deduced from the partition function; 
 for example the intrinsic entropy has the value
\begin{equation}
S=\frac{\beta^2}{16\pi}\{4-\frac{[1+(\frac {2m}{R_0})^2]
[3+2(\frac {2m}{R_0})^2] }{1-(\frac {2m}{R_0})^2}\},
\end{equation}
revealing a complex dependence on the parameters $m, R_0$. 
We are facing two well-known limits, namely, the Schwarzschild 
limit (a vanishing cosmological constant, $R_0 \to \infty$) 
with $S=4\pi m^2$ and the Einstein limit ($m \to 0$), with 
$S=0$ as expected. A curious result is obtained for a 
large enough $m$ ($\frac{m}{R_0} \gg 1 $); in this limit the 
asymptotic behavior of the entropy is $S\approx 8\pi m^2$.

However, in the $"extremal"$ case $r_H=\frac{\pi R_0}{4}$ ($i.e.$ $R_0=2m$)
 we obtain a diverging value for the entropy.
 This divergence is to be associated with the peculiar global 
 structure of the solution and the existence of the naked 
 singularity.
 
For $\frac{m}{R_0} \ll 1$ we obtain a set of corrections to the (asymptotically flat) 
Schwarzschild black hole thermodynamic quantities.
In this case, a curious result is the rather benign character of
 the antipodal naked singularity.

\section{Further discussions}
The exact solution proposed by Vaidya in \cite{6} is usually regarded as
a possible approach to the 
study of black holes in a non-flat background.

We have discussed in this paper some properties of this solution.
After considering the maximal analytic extension of this model,
 an interpretation has been proposed as describing a black hole with a 
antipodal naked singularity in the 
cosmological background of Einstein universe.
Assuming the possibility of extending path integral
 formulation of gravitational thermodynamics 
to this case, the expression of the Hawking temperature, the
Euclidean action and the intrinsic entropy have been derived.
These relations reveal a complex dependence on the two parameters of the
model $m, R_0$. 

The existence of a naked singularity in the line element (\ref{metric}) 
is certainly the most unpleasant feature of the Vaidya model.

For $\frac{m}{R_0} \ll 1$ 
a possibly way to deal with the naked singularity 
is to suppose that the line element (\ref{metric}) is 
valid for $r<r_0$ only (for a suitable $r_0>r_H$) and to match 
it to a different metric which goes over into the Einstein line-element for 
large enough $r$. 
In this way the Vaydia model becomes an isolated island 
in a Einstein universe background, 
and the problems associated with the case $R_0=2m$ can be avoided too.
A possible choice for $r_0$ is $r_0=\pi R_0/2$, where the line element (\ref{metric})
becomes the line element of the Einstein universe.
 Unfortunately, a surface density distribution
of matter seems to be always necessary.

A further extension of this paper could be the inclusion of a nonzero 
angular momentum and considering the immersion in 
an FRW universe along the line suggested by Vaidya in \cite{6}. 
Some features of the simplest case discussed in this paper 
(for example the existence of a naked singularity) are still 
present in the general case given the strong 
correlation between the particle and the universe.

 The Vaidya model represents an oversimplified picture and can not 
 be considered a live candidate for describing a physical
situation, but can be a source of insight into the possibilities 
allowed by relativity theory.
\\
\\
{\bf Acknowledgement}
\\
The authors gratefully acknowledge the referee's helpful comments.

\newpage
\begin{figure}
\begin{center}
\includegraphics[angle=270, width=1.2\textwidth]{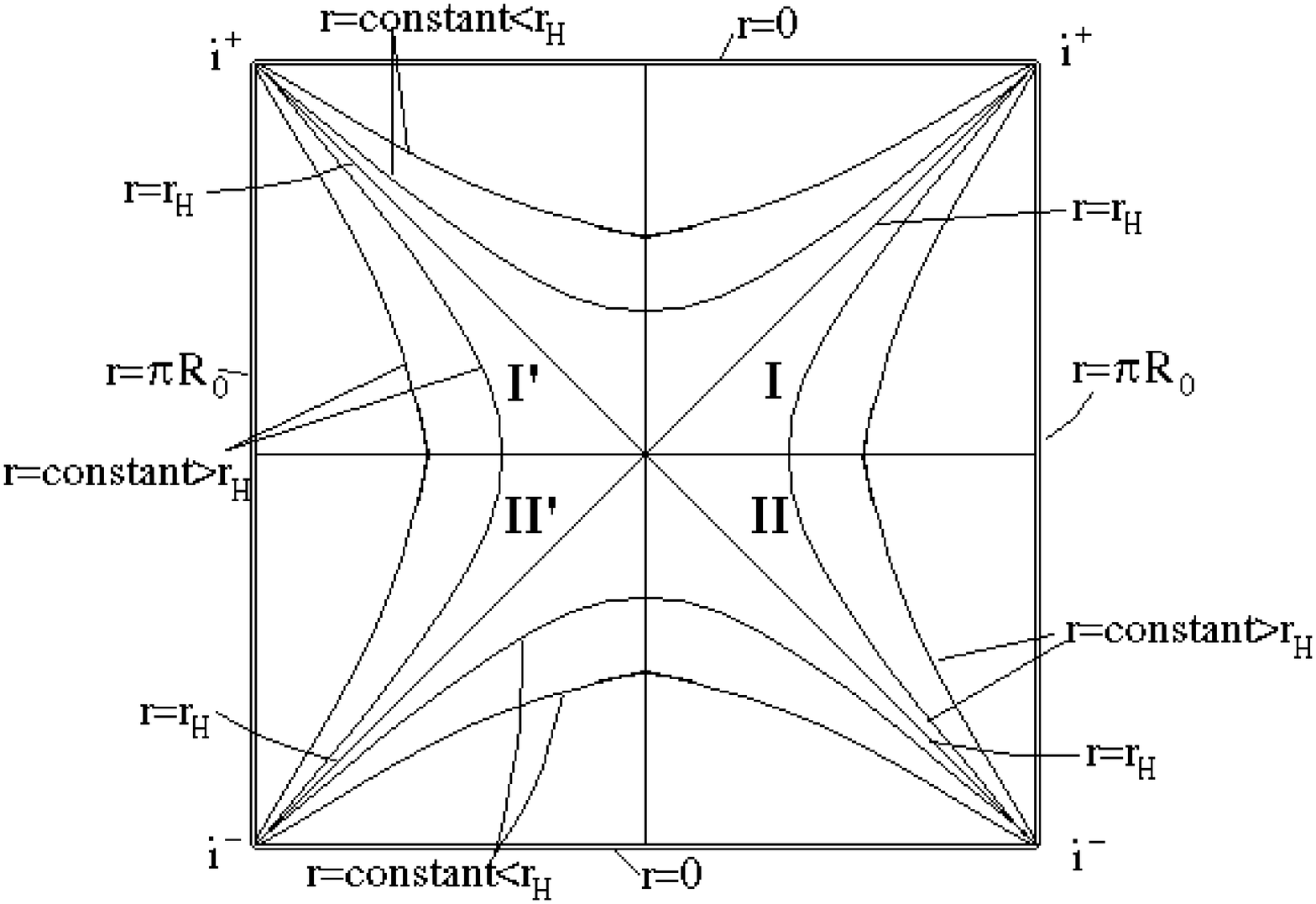}
\\[2cm]
\caption{Penrose diagram for the maximally extended Vaidya metric (\ref{metric})}
\end{center}
\end{figure}

\end{document}